\documentclass[conference]{IEEEtran}
\IEEEoverridecommandlockouts
\usepackage{cite}
\usepackage{amsmath,amssymb,amsfonts}
\usepackage{algorithmic}
\usepackage{graphicx}
\usepackage{textcomp}
\usepackage{xcolor}
\usepackage{quantikz} % for quantum circuit diagrams
\usepackage{braket}
\usepackage{adjustbox}
\def\BibTeX{{\rm B\kern-.05em{\sc i\kern-.025em b}\kern-.08em
    T\kern-.1667em\lower.7ex\hbox{E}\kern-.125emX}}
\begin{document}

\title{Digital Quantum Simulation of the quantum $\beta$-FPUT Lattice: Formulation and Resource Estimation  \\
% {\footnotesize \textsuperscript{*}Note: Sub-titles are not captured for https://ieeexplore.ieee.org  and
}
\author{\IEEEauthorblockN{1\textsuperscript{st} Kiratholly Nandakumar Madhav Sharma }

--------------------

\IEEEauthorblockA{\textit{LPMMC} \\
Univ. Grenoble Alpes, CNRS, 38000 Grenoble, France}

\IEEEauthorblockA{\textit{Capgemini Quantum Lab},\\
Place de l'Étoile, 11 rue de Tilsitt, 75017 Paris, France\\
0009-0000-0279-1086}

\and

\IEEEauthorblockN{2\textsuperscript{nd}  Juan Manuel Aguiar Hualde }

--------------------

\IEEEauthorblockA{\textit{Capgemini Quantum Lab},\\
Place de l'Étoile, 11 rue de Tilsitt, 75017 Paris, France\\
0000-0003-1370-5459}

\and

\IEEEauthorblockN{3\textsuperscript{rd} Julian Van Velzen}

--------------------

\IEEEauthorblockA{\textit{Capgemini Quantum Lab},\\
Place de l'Étoile, 11 rue de Tilsitt, 75017 Paris, France\\
0009-0000-6952-189X}

\and

\IEEEauthorblockN{ Phalgun Lolur$^*$}

--------------------

\IEEEauthorblockA{\textit{Capgemini Quantum Lab},\\
Place de l'Étoile, 11 rue de Tilsitt, 75017 Paris, France\\
$^*$ Corresponding author:phalgun.lolur@capgemini.com}
}

\maketitle

\begin{abstract}
% Heat conduction in low-dimensional systems is of broad fundamental and technological interest, especially in materials where anomalous transport and strong lattice nonlinearity play an important role.
% The $\beta$-Fermi–Pasta–Ulam–Tsingou ($\beta$-FPUT) chain provides a minimal nonlinear lattice model for investigating anomalous heat conduction and phonon relaxation, but its quantum real-time dynamics are difficult to access with conventional methods. 
% Here we develop a first-quantized digital quantum-simulation framework for the quantum $\beta$-FPUT model.
% We derive the quantized Hamiltonian, design Trotterized circuit blocks for time evolution, and propose a measurement protocol for mode-resolved correlation functions associated with phonon damping. 
% We also provide qubit, gate, and circuit-depth estimates relevant to fault-tolerant implementations. This work establishes a methodological basis for studying transport in nonlinear low-dimensional lattices on quantum computers.
Heat conduction in low-dimensional systems exhibits strong deviations from Fourier behavior due to anharmonicity and long-lived vibrational correlations, challenging conventional computational approaches. The $\beta$-Fermi--Pasta--Ulam--Tsingou ($\beta$-FPUT) chain provides a minimal nonlinear lattice model for studying anomalous transport, yet its quantum real-time dynamics remain difficult to access with classical methods. We develop a first-quantized digital quantum-simulation framework for the quantum $\beta$-FPUT lattice, targeting fault-tolerant quantum computers. By working directly with discretized lattice displacements rather than truncated phonon occupation spaces, the approach captures anharmonic interactions while avoiding bosonic encoding overheads. We construct Trotterized circuit blocks for real-time evolution and introduce a Hermitian quadrature decomposition of Fourier-mode displacement operators that enables shallow quantum circuits for mode-resolved displacement correlators. We analyze the quantum resources required for the full simulation and measurement workflow, providing qubit counts, gate complexity, circuit-depth and resource estimates as functions of system size and resolution within a fault-tolerant workflow. These results establish a concrete algorithmic blueprint for simulating quantum transport dynamics in nonlinear low-dimensional lattice models on fault-tolerant quantum hardware.

\end{abstract}


\section{Introduction}
\begin{figure}[h!]
\centering
\begin{tikzpicture}[font=\footnotesize]
\node[
draw=blue!50!black, fill=blue!8, rounded corners=5pt,
line width=0.9pt, align=center, text width=2cm,
minimum height=1.45cm, inner sep=2.5pt
] (A) at (0,1.25)
{\textbf{Electronics}\\
Hot spots, high heat flux};

\node[
draw=green!45!black, fill=green!8, rounded corners=5pt,
line width=0.9pt, align=center, text width=2cm,
minimum height=1.45cm, inner sep=2.5pt
] (B) at (0,-1.25)
{\textbf{Bio-devices}\\
Heating and biocompatibility};

\node[
draw=orange!65!black, fill=orange!10, rounded corners=6pt,
line width=1.0pt, align=center, text width=2.25cm,
minimum height=1.9cm, inner sep=3pt
] (C) at (3,0)
{\textbf{Shared need}\\
Efficient, controllable\\
heat dissipation};

\node[
draw=purple!60!black, fill=purple!8, rounded corners=6pt,
line width=1.0pt, align=center, text width=3cm,
minimum height=2.3cm, inner sep=3pt
] (D) at (6.25,0)
{\textbf{1D transport platforms}\\
Nanotubes, nanowires,\\
polymer chains (\emph{often biocompatible}),\\
and nonlinear lattice models};

\draw[-{Latex[length=3mm,width=2mm]}, line width=1.2pt, gray!75] (A.east) -- (C.west);
\draw[-{Latex[length=3mm,width=2mm]}, line width=1.2pt, gray!75] (B.east) -- (C.west);
\draw[-{Latex[length=3mm,width=2mm]}, line width=1.2pt, gray!75] (C.east) -- (D.west);
\end{tikzpicture}
\caption{Conceptual motivation for studying thermal transport in low-dimensional systems.}
\label{fig:motivation_schematic}
\end{figure}
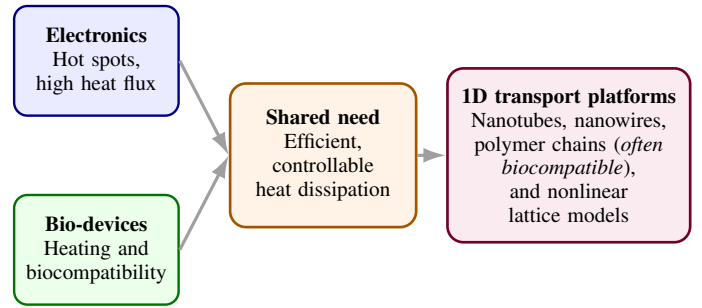

Thermal transport is a problem of both fundamental and technological importance. As summarized schematically in Fig.~\ref{fig:motivation_schematic}, distinct application domains share a common challenge: heat must be dissipated efficiently, locally, and under increasingly restrictive material constraints. In high-power microelectronics, continued device miniaturization and rising power densities produce severe hot spots; in systems such as GaN and Ga$_2$O$_3$ transistors, local heat fluxes can approach $300~\mathrm{kW/cm^2}$, leading to large thermal gradients, degraded performance, and reliability concerns~\cite{soleimanzadeh2019near}. Similar constraints arise in bio-implantable and flexible devices, where thermal dissipation must be controlled without compromising functionality or biocompatibility. Although heat spreaders based on Cu, Al, and near-junction diamond films can partially alleviate these issues, the broader need is for materials and predictive frameworks capable of describing efficient and controllable heat flow across both industrial and biomedical settings~\cite{soleimanzadeh2019near,xue2019thermal,xu2019nanostructured}.

This perspective naturally draws attention to low-dimensional transport platforms, including polymeric and other chain-like materials. Polymers are particularly appealing because they combine thermal functionality with mechanical flexibility and, in some cases, biocompatibility~\cite{xue2019thermal,xu2019nanostructured}. Although polymers are often regarded as thermal insulators, their conductivity can increase dramatically when chains are highly aligned. Xu \emph{et al.} reported thermal conductivities of $62~\mathrm{W\,m^{-1}K^{-1}}$ in nanostructured polyethylene films, while Henry and Chen found from molecular-dynamics simulations that an individual polyethylene chain can exhibit divergent conductivity in the quasi-one-dimensional limit~\cite{xu2019nanostructured,henry2008high}. Together, these observations identify polymeric and other one-dimensional systems as a natural setting in which application-driven thermal management meets the fundamental physics of anomalous heat transport.

Understanding such behavior requires predictive microscopic theories beyond conventional diffusive descriptions. Fourier-based frameworks successfully describe ordinary heat conduction but fail to capture ballistic and non-Fourier transport at micro- and nanoscales~\cite{johnson2015non}. As a result, nonlinear lattice-dynamical models have become central tools for investigating anomalous transport in low-dimensional systems. Their relevance connects directly to the Fermi--Pasta--Ulam--Tsingou (FPUT) problem, originally introduced to study thermalization in nonlinear chains. Since then, FPUT-type models have served as paradigmatic frameworks for energy equipartition, phonon relaxation, and size-dependent heat transport in one-dimensional systems~\cite{Wang_2024,Hou_2012}. These behaviors closely parallel the anomalous transport observed in polymer chains and related quasi-one-dimensional materials~\cite{henry2008high}.

A variety of computational approaches have been employed to study heat transport in $\beta$--FPUT-like systems, each capturing complementary but incomplete aspects of the problem, as summarized in Table~\ref{tab:method_comparison}. Classical molecular dynamics provides direct access to real-time dynamics and remains a powerful tool for studying anomalous transport, but it neglects quantum statistics, zero-point motion, and operator-valued observables. Path Integral Monte Carlo (PIMC), by contrast, accurately captures equilibrium quantum properties by mapping the system to imaginary time, but real-time observables must then be recovered through analytic continuation. This procedure is inherently ill-posed, with small statistical uncertainties often amplified exponentially, making phonon lifetimes and transport coefficients difficult to extract reliably~\cite{jarrell1996bayesian,ceperley1995path}. These limitations are particularly severe in low-dimensional systems, where vibrational quantization, zero-point motion, and long phonon mean free paths strongly influence transport~\cite{togo2015first}.

\begin{table}[t]
\centering
\begin{tabular}{p{2cm} p{2.5cm} p{2.4cm}}
\hline
\textbf{Method} & \textbf{Main strength} & \textbf{Main limitation for the present problem} \\
\hline
Molecular dynamics (MD) & Direct access to real-time dynamics; efficient for large classical systems & Neglects quantum statistics, zero-point motion, and operator dynamics \\
\hline
Path Integral Monte Carlo (PIMC) & Accurate equilibrium quantum properties in imaginary time & Real-time observables require analytic continuation, which is ill-posed and unreliable for transport coefficients \\
\hline
Digital quantum simulation & Direct unitary evolution of the quantized lattice Hamiltonian; access to real-time quantum correlators & Large-scale simulations require deep circuits and fault-tolerant hardware \\
\hline
\end{tabular}
\caption{Comparison of approaches for studying transport in low-dimensional lattice models.}
\label{tab:method_comparison}
\end{table}

These considerations motivate the study of the quantum $\beta$--FPUT model, namely the quantized extension of the classical FPUT chain widely used in statistical mechanics. In the quantum framework, lattice displacements $q_j$ and conjugate momenta $p_j$ are promoted to operators obeying canonical commutation relations, and the dynamics are generated by unitary time evolution under the corresponding Hamiltonian. This formulation provides direct access to quantum dynamical observables, including mode-resolved correlation functions such as $\langle \hat Q_k(t)\hat Q_k(0) \rangle$, which are directly relevant to phonon lifetimes and anomalous transport in low-dimensional quantum materials.

The approach explored in this work is explicitly designed for fault-tolerant quantum computing and lies beyond the capabilities of noisy intermediate-scale quantum (NISQ) hardware. Recent studies of vibrational Hamiltonians indicate that small systems may be benchmarked on classical emulators or present-day devices, whereas accurate real-time simulations at larger system sizes require deep Trotterized circuits and are most naturally framed in a fault-tolerant setting~\cite{kuanysheva2025quantum,miessen2021quantum,malpathak2025trotter}.

As a part of this study, we develop a first-quantized digital quantum-simulation framework for the quantum $\beta$--FPUT chain. Starting from the real-space lattice Hamiltonian, we derive its quantum formulation, construct Trotterized circuit blocks for real-time evolution, and introduce a protocol for extracting mode-resolved displacement correlators. 
A central element of the approach is a Hermitian quadrature decomposition of Fourier-mode displacement operators, which enables shallow quantum circuits for correlation-function estimation. 
We further provide qubit, gate, and circuit-depth estimates relevant to fault-tolerant implementations, framing the full simulation and measurement procedue as an end-to-end execution workflow suitable for fault-tolerant quantum systems. 
Our aim is to establish a concrete algorithmic foundation for quantum simulations of transport in nonlinear low-dimensional lattice models.

The remainder of this article is organized as follows. Section~\ref{sec:Mathematical_Formulation} presents the mathematical formulation of the quantum $\beta$--FPUT model. Section~\ref{Unitary_for_hamiltonian_evolution} describes the construction of unitary evolution operators, while Section~\ref{sec:Fourier_Mode_Displacement_Operator} introduces the protocol for extracting mode-resolved correlators. Resource estimates are discussed in Section~\ref{sec:Resource_Estimation}, and Section~\ref{sec:Conclusion_Outlook} concludes with a summary and outlook.

\section{Mathematical Formulation of the Quantum FPUT Model}\label{sec:Mathematical_Formulation}

We consider the $\beta$--Fermi--Pasta--Ulam--Tsingou (FPUT) chain: a one-dimensional lattice of $N$ identical masses with nearest-neighbor interactions. In real space, the classical Hamiltonian is given by
\begin{equation}
H =
\sum_{j=1}^N \frac{p_j^2}{2m}
+ \frac{\kappa}{2} \sum_{j=1}^N (q_{j+1}-q_j)^2
+ \frac{\beta}{4} \sum_{j=1}^N (q_{j+1}-q_j)^4 ,
\label{eq:FPUT_Hamiltonian}
\end{equation}
where $q_j$ denotes the displacement of site $j$, $p_j$ its conjugate momentum, $m$ the mass, $\kappa$ the harmonic coupling, and $\beta$ the quartic anharmonicity. The quartic term governs phonon--phonon scattering and is the microscopic origin of relaxation and anomalous transport in one-dimensional momentum-conserving lattices.

\begin{figure}[h]
\centering
\includegraphics[width=\linewidth]{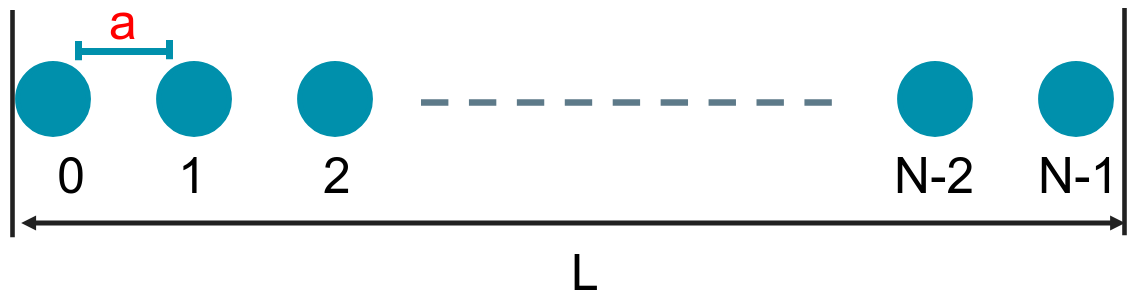}
\caption{Schematic diagram of the one-dimensional FPUT chain with sites labeled $\{0,1,2,\ldots,N-1\}$.}
\label{fig:FPU_Diagram}
\end{figure}

Figure~\ref{fig:FPU_Diagram} illustrates the real-space geometry of the chain with periodic boundary conditions, which are assumed throughout this work.

For analytical insight, it is convenient to transform to Fourier space. Defining the discrete Fourier transforms
\begin{align}
Q_k &= \frac{1}{\sqrt{N}} \sum_{j=1}^N q_j \, e^{-i \tfrac{2\pi}{N} j k}, &
P_k &= \frac{1}{\sqrt{N}} \sum_{j=1}^N p_j \, e^{-i \tfrac{2\pi}{N} j k},
\end{align}
the canonical structure becomes $\{Q_k,P_{k'}\}=\delta_{k,k'}$. In these variables, the harmonic part of the Hamiltonian diagonalizes as
\begin{equation}
H_{\text{harm}} =
\sum_{k} \left(
\frac{P_k^2}{2m} + \frac{m \omega_k^2}{2} Q_k^2
\right),
\end{equation}
with dispersion relation
\begin{equation}
\omega_k^2 = \frac{2\kappa}{m} \bigl(1 - \cos(ka) \bigr),
\end{equation}
where $a$ is the lattice spacing. The quartic interaction becomes a momentum-conserving four-mode coupling of the form
\begin{equation}
H_{\text{quartic}} =
\sum_{k_1,k_2,k_3,k_4}
T_{k_1,k_2,k_3,k_4}\,
Q_{k_1} Q_{k_2} Q_{k_3} Q_{k_4}\,
\delta_{k_1+k_2+k_3+k_4,\,0}.
\end{equation}

We now pass to the quantum model by canonical quantization, promoting the classical variables to operators satisfying
\begin{equation}
[\hat q_j,\hat p_{j'}] = i\hbar\,\delta_{j,j'},
\end{equation}
or equivalently $[\hat Q_k,\hat P_{k'}]=i\hbar\,\delta_{k,k'}$ in Fourier space. The quantum Hamiltonian then reads
\begin{align}
\hat H &=
\sum_{k} \left(
\frac{\hat P_k^2}{2m} + \frac{m \omega_k^2}{2} \hat Q_k^2
\right) \nonumber \\
&\quad +
\sum_{k_1,k_2,k_3,k_4}
T_{k_1,k_2,k_3,k_4}\,
\hat Q_{k_1} \hat Q_{k_2} \hat Q_{k_3} \hat Q_{k_4}\,
\delta_{k_1+k_2+k_3+k_4,\,0}.
\end{align}

A commonly used representation introduces phonon creation and annihilation operators
\begin{align}
\hat a_k &=
\sqrt{\frac{m \omega_k}{2\hbar}}\, \hat Q_k
+ \frac{i}{\sqrt{2 m \hbar \omega_k}} \hat P_{-k}, \\
\hat a_k^\dagger &=
\sqrt{\frac{m \omega_k}{2\hbar}}\, \hat Q_{-k}
- \frac{i}{\sqrt{2 m \hbar \omega_k}} \hat P_k ,
\end{align}
which satisfy $[\hat a_k,\hat a_{k'}^\dagger]=\delta_{k,k'}$ and yield the diagonal harmonic Hamiltonian
\begin{equation}
\hat H_{\text{harm}} =
\sum_k \hbar \omega_k \left(\hat a_k^\dagger \hat a_k + \tfrac{1}{2}\right).
\end{equation}
The quartic term generates mode couplings that can be written schematically as
\begin{equation}
\hat V_4 =
\sum_{k_1,k_2,k_3,k_4}
U_{k_1,k_2,k_3,k_4}\,
\hat a_{k_1}^\dagger \hat a_{k_2}^\dagger
\hat a_{k_3} \hat a_{k_4}\,
\delta_{k_1+k_2,\,k_3+k_4},
\end{equation}
so that $\hat H=\hat H_{\text{harm}}+\hat V_4$.

While this phonon representation is physically transparent, it is not the most convenient starting point for a digital quantum simulation. Bosonic modes have infinite local dimension, and truncation at a maximum occupation introduces a systematic approximation that becomes increasingly costly with both system size and anharmonicity. Moreover, because the interaction originates from quartic displacements, the resulting operator structure contains phonon-number nonconserving terms beyond simple $2\leftrightarrow2$ scattering, complicating circuit synthesis. In addition, the quartic interaction couples many mode combinations, leading to dense operator support and potentially unfavorable product-formula constants. Qubit encodings of truncated bosonic spaces further introduce significant overheads in both qubit count and gate depth.

Motivated by these considerations, we adopt a first-quantized formulation that retains the natural separation between kinetic and potential energy,
\begin{align}
\hat H &= \sum_{j=1}^N \frac{\hat p_j^2}{2m} + V(\mathbf{q}), \\
V(\mathbf{q}) &=
\frac{\kappa}{2} \sum_j (q_{j+1}-q_j)^2
+ \frac{\beta}{4} \sum_j (q_{j+1}-q_j)^4 .
\end{align}
This representation is directly compatible with split-operator (Trotter--Suzuki) time evolution: the kinetic term is diagonal in the momentum basis, while the potential term is diagonal in the position basis. Continuous variables are discretized on finite grids with controlled resolution error, avoiding explicit occupation-number truncations. The remainder of this work therefore focuses on implementing the first-quantized unitary time-evolution operator and estimating the quantum resources required for its execution.

\section{Implementation of the Unitary Evolution Operator $(U(t)=e^{-iHt/\hbar})$}\label{Unitary_for_hamiltonian_evolution}
The real-time dynamics of the quantum FPUT lattice are governed by the time-dependent Schr\"odinger equation
\begin{equation}
i \hbar \frac{d}{dt} \ket{\psi(t)} = H \ket{\psi(t)} ,
\end{equation}
where $H$ is the system Hamiltonian. The formal solution is given by the unitary propagator
\begin{equation}
U(t) = e^{- i H t / \hbar},
\label{eq:Unitary_Evolution}
\end{equation}
so that $\ket{\psi(t)} = U(t)\ket{\psi(0)}$. For lattice observables such as the displacement operator $\hat{Q}_j$,
the Heisenberg-picture evolution reads
\begin{equation}
\hat{Q}_j(t) = U^\dagger(t)\,\hat{Q}_j(0)\,U(t),
\end{equation}
which enables the evaluation of dynamical correlation functions
\begin{equation}
C_{jk}(t) = \langle \psi(0) |\, \hat{Q}_j(t)\,\hat{Q}_k(0)\,| \psi(0) \rangle .
\end{equation}

Since the kinetic and potential contributions of the Hamiltonian do not commute, we approximate $U(t)$ using a
Trotter--Suzuki product formula. A first-order (Lie--Trotter) decomposition,
\begin{equation}
U(t) \approx \left( e^{- i H_{\text{kin}} \Delta t / \hbar}\; e^{- i H_{\text{pot}} \Delta t / \hbar} \right)^n ,
\qquad n=t/\Delta t,
\end{equation}
introduces a local error of order $\mathcal{O}\!\left(\Delta t^{2}\|[H_{\text{kin}},H_{\text{pot}}]\|\right)$ and therefore a
global error $\varepsilon_{\rm tot}^{(1)}=\mathcal{O}\!\left(t\,\Delta t\,\|[H_{\text{kin}},H_{\text{pot}}]\|\right)$.
To reduce this error at fixed circuit depth per time step, we employ the standard symmetric second-order splitting
\cite{nielsen2010quantum,malpathak2025trotter,miessen2021quantum},
\begin{equation}
\begin{aligned}
U(t) \approx {}&
\left(
e^{- i H_{\text{kin}} \frac{\Delta t}{2\hbar}}
\, e^{- i H_{\text{pot}} \frac{\Delta t}{\hbar}}
\, e^{- i H_{\text{kin}} \frac{\Delta t}{2\hbar}}
\right)^n
\\[3pt]
={}&
\left(
U_{\text{kin}}\!\left(\tfrac{\Delta t}{2}\right)\,
U_{\text{pot}}(\Delta t)\,
U_{\text{kin}}\!\left(\tfrac{\Delta t}{2}\right)
\right)^n .
\end{aligned}
\label{eq:Trotterization_implemented}
\end{equation}
For this symmetric second-order formula, the leading error term is governed by nested commutators and scales as
\begin{equation}
\epsilon_{\text{Trotter}} = \mathcal{O}\!\left( \Delta t^{3} \, \| [H_{\text{kin}},[H_{\text{kin}},H_{\text{pot}}]]
+ [H_{\text{pot}},[H_{\text{pot}},H_{\text{kin}}]] \| \right),
\end{equation}
so that after $n=t/\Delta t$ time slices the accumulated error scales as
$\varepsilon_{\rm tot}^{(2)}=\mathcal{O}(t\,\Delta t^{2})$ up to the corresponding operator-norm prefactor.

This scaling highlights the trade-off between first- and higher-order product formulas. For a fixed target accuracy
$\varepsilon$, the step count improves from $n=\mathcal{O}(t^{2}/\varepsilon)$ for first order to
$n=\mathcal{O}(t^{3/2}/\sqrt{\varepsilon})$ for the symmetric second-order scheme. More generally, a $2p$-th order Suzuki
formula yields $\varepsilon_{\rm tot}^{(2p)}=\mathcal{O}(t\,\Delta t^{2p}\Lambda_{2p})$, where $\Lambda_{2p}$ collects
$(2p)$-fold nested commutators, implying
\begin{equation}
n=\mathcal{O}\!\left(t^{1+1/(2p)}\,\Lambda_{2p}^{1/(2p)}\,\varepsilon^{-1/(2p)}\right).
\label{eq:Number_of_time_steps_for_2p-th_order_suzuki}
\end{equation}
The improved $\varepsilon$-dependence is obtained at the expense of a scheme-dependent constant-factor increase in the number
of exponentials (and thus circuit depth) per time slice, since higher-order constructions contain more factors of
$e^{-iH_{\text{kin}}\,\cdot/\hbar}$ and $e^{-iH_{\text{pot}}\,\cdot/\hbar}$. In the remainder of this section, we describe how
each exponential factor is realized as a unitary circuit block, ensuring that the overall time-evolution operator remains
unitary by construction.

\subsection{Implementation of the Kinetic Operator}\label{sec:Kinetic_operator_Implementation}

The kinetic contribution to the Hamiltonian is
\begin{equation}
H_{\text{kin}}=\sum_{j=1}^{N}\frac{\hat p_j^{2}}{2m},
\label{eq:H_kinetic}
\end{equation}
where $\hat p_j$ is the momentum operator at site $j$ and $m$ is the particle mass. Since $H_{\text{kin}}$ is quadratic in momentum, it is diagonal in the momentum basis, which makes its real-time propagator particularly amenable to a digital implementation via phase rotations. For the symmetric Trotter--Suzuki decomposition used in this work, we require the kinetic evolution over a half time-step,
\begin{equation}
U_{\text{kin}}\!\left(\tfrac{\Delta t}{2}\right)
=\exp\!\left(-\frac{i\Delta t}{2\hbar}\sum_{j=1}^{N}\frac{\hat p_j^{2}}{2m}\right)
=\prod_{j=1}^{N}\exp\!\left(-i\,\frac{\Delta t}{4m\hbar}\,\hat p_j^{2}\right),
\end{equation}
so that each momentum eigenstate $|p_j\rangle$ acquires the phase $\exp\!\left(-i\,p_j^{2}\Delta t/(4m\hbar)\right)$, encoding the free-particle dispersion $E(p)=p^{2}/(2m)$.
We represent each site coordinate $q_j$ on an $n$-qubit register $s_j$, i.e., the local basis is $\{|x\rangle:x\in\{0,1,\ldots,2^{n}-1\}\}$ and the full data register comprises $Nn$ qubits. Because the kinetic operator is diagonal in momentum but our potential operator is most naturally applied in the position basis, we implement $U_{\text{kin}}(\Delta t/2)$ by the standard basis-change pattern \emph{QFT--phase--inverse--QFT} applied independently to each site register: an $n$-qubit quantum Fourier transform $\mathrm{QFT}_n$ maps $|q_j\rangle$ to a discrete momentum basis $|p_j\rangle$, a diagonal operator $D_p(\Delta t/2)$ applies the required momentum-dependent phase, and $\mathrm{QFT}_n^\dagger$ returns to the position representation.
This sequence reproduces the desired kinetic half-step exactly (up to the discretization of the continuous variables) while keeping the circuit compatible with a real-space implementation of the potential propagator.

The corresponding circuit structure is shown in Fig.~\ref{fig:kinetic}. E
ach $\mathrm{QFT}_n$ block can be compiled into one-qubit Hadamards and two-qubit controlled-phase rotations $cR_\ell$ (exactly, requiring $\mathcal{O}(n^{2})$ two-qubit gates, approximately, by truncating small-angle rotations) \cite{nielsen2010quantum}.
The diagonal block $D_p(\Delta t/2)$ can be realized using phase kickback: in the QFT basis one encodes $p_j$ in its binary representation, reversibly computes $p_j^{2}$ into an ancilla work register, applies a controlled $R_z$ (or an equivalent phase-gradient operation) with rotation angle proportional to $p_j^{2}\Delta t/(4m\hbar)$, and finally uncomputes the work register, ensuring reversibility and leaving only the desired phase on the data register. 
This implementation is strictly diagonal in the momentum basis and therefore preserves unitarity by construction. 
Resource-wise, the data register requires $Nn$ qubits, while the phase-kickback implementation of $D_p$ uses an additional $a$ ancilla qubits for reversible arithmetic (with $a=\mathcal{O}(n)$--$\mathcal{O}(n\log n)$ depending on the chosen squaring circuit); these ancillas may be reused serially across sites (total $Nn+a$ qubits) or replicated for full parallelism (total $N(n+a)$). 
For depth estimates, an exact $\mathrm{QFT}_n$ has depth $\mathcal{O}(n^{2})$ in standard decompositions \cite{nielsen2010quantum}, and $D_p$ is dominated by reversible squaring and controlled-phase application, which also scales as $\mathcal{O}(n^{2})$ in Toffoli/controlled-rotation depth for elementary arithmetic. Therefore,
\begin{equation}
\begin{aligned}
\mathrm{Depth}\!\left[
U_{\mathrm{kin}}\!\left(\tfrac{\Delta t}{2}\right)
\right]
={}& \mathcal{O}\!\Bigl(
      \mathrm{Depth}(\mathrm{QFT}_n)
      + \mathrm{Depth}(D_p)
\\
&\qquad\quad
      + \mathrm{Depth}(\mathrm{QFT}_n^\dagger)
\Bigr)
\\
={}& \mathcal{O}(n^2).
\end{aligned}
\end{equation}
up to compilation-dependent constants. Table~\ref{tab:kinetic_resources} reports illustrative qubit counts for representative system sizes under serial ancilla reuse and fully parallel application of $D_p$, using $n=6$ (a 64-point grid per site) and $a=2n$ as a conservative arithmetic-work budget.

\begin{figure}[h!]
\centering
\footnotesize
\begin{adjustbox}{max width=\columnwidth}
\begin{quantikz}[row sep=0.35cm, column sep=0.35cm]
\lstick{$s_0:\;n~\text{qubits}$} &
\gate{\begin{array}{c}
\mathrm{QFT}_n\\[-1mm]
\scriptstyle (H+cR_\ell)\\[-1mm]
\scriptstyle \sim \mathcal{O}(n^2)\;2q
\end{array}} &
\gate{\shortstack{$D_p(\Delta t/2)$\\[-1mm]\scriptsize phase kickback\\[-1mm]\scriptsize $p^2$ via rev.\ arith.}} &
\gate{\mathrm{QFT}_n^\dagger} & \qw \\
\lstick{\vdots} &
\gate{\mathrm{QFT}_n} &
\gate{D_p(\Delta t/2)} &
\gate{\mathrm{QFT}_n^\dagger} & \qw \\
\lstick{$s_{N-1}:\;n~\text{qubits}$} &
\gate{\mathrm{QFT}_n} &
\gate{D_p(\Delta t/2)} &
\gate{\mathrm{QFT}_n^\dagger} & \qw
\end{quantikz}
\end{adjustbox}

\vspace{0.3em}
\begin{minipage}{\columnwidth}
\footnotesize
\textbf{Registers:} Data registers are $N$ site registers $s_j$ of size $n$ qubits each ($Nn$ qubits total). Optional ancillas: $a$ qubits for reversible evaluation of $p_j^2$, reused serially or replicated for parallel execution.
\end{minipage}

\caption{Kinetic half-step circuit. Each $n$-qubit $\mathrm{QFT}_n$ is decomposed into Hadamard and controlled-phase gates $cR_\ell$. The exact QFT requires $\mathcal{O}(n^2)$ two-qubit gates, while approximate QFT reduces the count by truncating small-angle rotations. The diagonal operator $D_p(\Delta t/2)$ applies the phase $e^{-i p_j^2 \Delta t/(4m\hbar)}$ in the momentum basis via phase kickback using reversible arithmetic followed by uncomputation.}
\label{fig:kinetic}
\end{figure}

\begin{table}[t]
\centering
\footnotesize
\renewcommand{\arraystretch}{1.15}
\setlength{\tabcolsep}{3pt}
\label{tab:kinetic_resources}
\begin{tabular}{p{0.12\columnwidth} p{0.14\columnwidth} p{0.31\columnwidth} p{0.31\columnwidth}}
\hline
\centering System size $N$ &
\centering Grid bits/site $n$ &
\centering Qubits (serial ancilla reuse) $Nn+a$ &
\centering Qubits (fully parallel) $N(n+a)$
\tabularnewline
\hline
\centering $8$  & \centering $6$ & \centering $8\cdot 6 + 12 = 60$  & \centering $8(6+12)=144$ \tabularnewline
\centering $16$ & \centering $6$ & \centering $16\cdot 6 + 12 = 108$ & \centering $16(6+12)=288$ \tabularnewline
\centering $32$ & \centering $6$ & \centering $32\cdot 6 + 12 = 204$ & \centering $32(6+12)=576$ \tabularnewline
\hline
\end{tabular}
\caption{Illustrative qubit requirements for implementing the kinetic half-step, showing the trade-off between serial ancilla reuse and full parallelism.
Example parameters: $n=6$ qubits per site (64-point grid) and ancilla budget $a=2n=12$ for reversible squaring workspace in $D_p$.}
\end{table}

\subsection{Implementation of the Potential Operator}\label{sec:Implementation_of_the_potential_operator}

The potential Hamiltonian is
\begin{equation}
H_{\text{pot}} = \sum_{j=1}^N \left[ \frac{\kappa}{2}(q_{j+1}-q_j)^2 + \frac{\beta}{4}(q_{j+1}-q_j)^4 \right],
\end{equation}
which is diagonal in the position basis. Its propagator is therefore expressed as
\begin{equation}
U_{\text{pot}}(\Delta t) = \exp\!\left( -\frac{i \Delta t}{\hbar} H_{\text{pot}} \right).
\end{equation}
To implement this operator, the key step is to construct the bond variables
\begin{equation}
\Delta_j = q_{j+1} - q_j ,
\end{equation}
which encode relative displacements between neighboring sites.
Once these relative coordinates are obtained, the quadratic and quartic contributions are simulated as controlled phase rotations conditioned on the digital representation of $\Delta_j$. 
The quadratic term produces two-body interactions, corresponding to harmonic coupling, while the quartic term generates higher-order multi-controlled phases representing anharmonicity.
To ensure translational invariance and efficient depth, the system is divided into two disjoint layers of bonds (even and odd), so that non-overlapping interactions can be applied in parallel. 
After the phases are applied, the auxiliary registers holding $\Delta_j$ are uncomputed, returning them to their initial state and leaving only the physical degrees of freedom entangled. 
This procedure guarantees that the potential propagator is implemented exactly in the digital encoding.

\bigskip

Combining the above implementation of the kinetic and potential operators, the complete Trotter step is realized as outlined in Eq.~\eqref{eq:Trotterization_implemented}, which maintains the unitary structure of the exact propagator while enabling efficient decomposition into quantum gates. 
This decomposition leverages the Fourier diagonalization of the kinetic energy and modular arithmetic for relative displacements in the potential, ensuring a resource-efficient quantum simulation of the $\beta$--FPUT lattice model  \cite{nielsen2010quantum,malpathak2025trotter,miessen2021quantum}.

\begin{figure}[t]
\centering
\footnotesize
\begin{quantikz}[row sep=0.42cm, column sep=0.28cm]
\lstick{$\cdots$} & \qw & \qw & \qw & \qw & \qw \\
\lstick{$s_0$}
& \gate[wires=3]{\scriptsize SUB}
& \qw
& \qw
& \gate[wires=3]{\scriptsize UNSUB}
& \qw \\
\lstick{$s_1$}
& \qw
& \qw
& \qw
& \qw
& \qw \\
\lstick{$\text{\scriptsize anc.}$}
& \qw
& \ctrl{1}
& \ctrl{1}
& \qw
& \qw \\
\lstick{$r_0$}
& \qw
& \gate{\Phi_2}
& \gate{\Phi_4}
& \qw
& \qw \\
\lstick{$\cdots$} & \qw & \qw & \qw & \qw & \qw
\end{quantikz}
\caption{Two-site potential-layer skeleton implementing subtraction, controlled phase kickback, and inverse subtraction. Here, SUB computes the relative displacement $\Delta_0=s_1-s_0$ into the ancilla register without disturbing $s_0$ and $s_1$, while UNSUB uncomputes it. The controlled phase gates are $\Phi_2=\exp[-i(\kappa/2)\Delta_0^2\Delta t/\hbar]$ and $\Phi_4=\exp[-i(\beta/4)\Delta_0^4\Delta t/\hbar]$, acting on the reference qubit $r_0$ and controlled by the ancilla.}
\label{fig:potential_layer}
\end{figure}
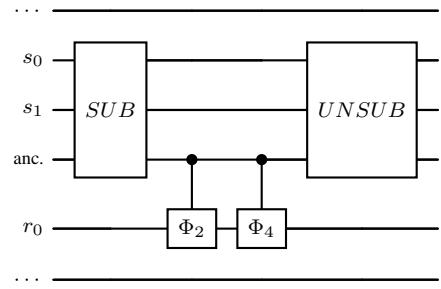

% \section{Implementation of the Unitary Evolution Operator $(U(t)=e^{-iHt/\hbar})$}\label{Unitary_for_hamiltonian_evolution}
\section{Implementation of Fourier--Mode Displacement Operators}\label{sec:Fourier_Mode_Displacement_Operator}

The Fourier transform of lattice displacements introduces a subtlety in quantum simulation. The mode operator
\begin{equation}
\hat Q_k = \frac{1}{\sqrt{N}} \sum_{j=0}^{N-1} \hat q_j \, e^{-i 2\pi j k / N}
\label{eq:Qk_def}
\end{equation}
is not Hermitian, even though each site displacement operator $\hat q_j$ is Hermitian and mutually commuting,
$[\hat q_j,\hat q_\ell]=0$. As a result, the exponential operator $e^{i\theta \hat Q_k}$ cannot be implemented directly as a unitary quantum circuit.

To address this issue, we reformulate $\hat Q_k$ in terms of Hermitian components, show how the corresponding unitaries can be implemented efficiently, and describe how mode-resolved displacement correlators are reconstructed from their expectation values.

\subsection{Hermitian Quadratures and Circuit Synthesis}

We decompose $\hat Q_k$ into its real and imaginary parts,
\begin{equation}
\hat Q_k = \hat Q_k^{(c)} - i \hat Q_k^{(s)},
\label{eq:Qk_decomp}
\end{equation}
with
\begin{align}
\hat Q_k^{(c)} &= \frac{1}{\sqrt{N}} \sum_{j=0}^{N-1}
\hat q_j \cos\!\left(\tfrac{2\pi j k}{N}\right),
\label{eq:Qk_cos} \\
\hat Q_k^{(s)} &= \frac{1}{\sqrt{N}} \sum_{j=0}^{N-1}
\hat q_j \sin\!\left(\tfrac{2\pi j k}{N}\right).
\label{eq:Qk_sin}
\end{align}
The operators $\hat Q_k^{(c)}$ and $\hat Q_k^{(s)}$ correspond to real standing-wave quadratures of the lattice displacement field. The cosine component $\hat Q_k^{(c)}$ projects onto the spatial pattern $\cos(2\pi j k/N)$, while the sine component $\hat Q_k^{(s)}$ projects onto $\sin(2\pi j k/N)$. Together, these two quadratures encode the amplitude and phase of the traveling-wave mode at wave number $k$.

Both $\hat Q_k^{(c)}$ and $\hat Q_k^{(s)}$ are Hermitian, and because they are real linear combinations of commuting position operators, they commute with each other,
\begin{equation}
[\hat Q_k^{(c)}, \hat Q_k^{(s)}] = 0.
\label{eq:commute}
\end{equation}
This property plays a central role in the measurement protocol developed below.

Because only Hermitian operators generate unitary evolutions of the form $e^{i\theta A}$ with real $\theta$, exponentials of $\hat Q_k$ must be reconstructed indirectly. The correlation function of interest can be expressed in terms of the Hermitian quadratures as
\begin{align}
\langle \hat Q_k(t)\hat Q_k(0) \rangle
&=
\langle \hat Q_k^{(c)}(t)\hat Q_k^{(c)}(0) \rangle
-
\langle \hat Q_k^{(s)}(t)\hat Q_k^{(s)}(0) \rangle
\nonumber \\
&\quad
- i\Big(
\langle \hat Q_k^{(c)}(t)\hat Q_k^{(s)}(0) \rangle
+
\langle \hat Q_k^{(s)}(t)\hat Q_k^{(c)}(0) \rangle
\Big).
\label{eq:Qk_reconstruct}
\end{align}
For equilibrium states with inversion symmetry, or more generally symmetry under $k\to -k$, the mixed quadrature correlators vanish, and the correlator becomes purely real up to finite-size and statistical effects.

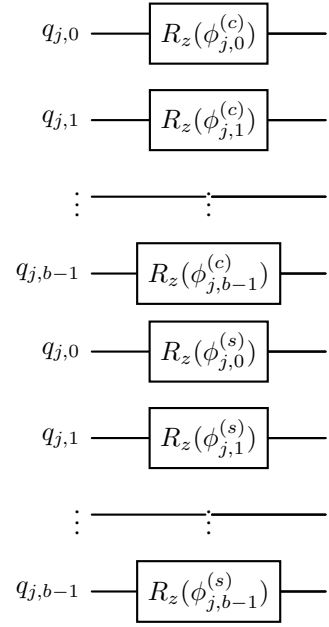
\begin{figure}[t]
\centering
\begin{quantikz}[row sep=0.35cm, column sep=0.6cm]
\lstick{$q_{j,0}$} & \gate{R_z(\phi^{(c)}_{j,0})} & \qw \\
\lstick{$q_{j,1}$} & \gate{R_z(\phi^{(c)}_{j,1})} & \qw \\
\lstick{$\vdots$}  & \vdots                       & \qw \\
\lstick{$q_{j,b-1}$} & \gate{R_z(\phi^{(c)}_{j,b-1})} & \qw
\end{quantikz}
\quad
\begin{quantikz}[row sep=0.35cm, column sep=0.6cm]
\lstick{$q_{j,0}$} & \gate{R_z(\phi^{(s)}_{j,0})} & \qw \\
\lstick{$q_{j,1}$} & \gate{R_z(\phi^{(s)}_{j,1})} & \qw \\
\lstick{$\vdots$}  & \vdots                       & \qw \\
\lstick{$q_{j,b-1}$} & \gate{R_z(\phi^{(s)}_{j,b-1})} & \qw
\end{quantikz}
\caption{Quantum circuits implementing $e^{i\theta \hat Q_k^{(c)}}$ (left) and $e^{i\theta \hat Q_k^{(s)}}$ (right). Each qubit $q_{j,r}$ is rotated by an angle $\phi^{(c)}_{j,r}$ or $\phi^{(s)}_{j,r}$ from Eq.~\eqref{eq:angles}. Since all rotations commute, the operations can be parallelized, producing depth-two circuits.}
\label{fig:Qk_circuits}
\end{figure}

Each site displacement operator $\hat q_j$ is diagonal in the position basis and is encoded on a $b$-qubit register with grid spacing $\Delta_q$,
\begin{equation}
\hat q_j =
\Delta_q \sum_{r=0}^{b-1} w_r^{\mathrm{bit}}
\, |1\rangle\!\langle 1|_{j,r},
\label{eq:qj_binary}
\end{equation}
where $w_r^{\mathrm{bit}}=2^r$ for unsigned encodings or $w_{b-1}^{\mathrm{bit}}=-2^{b-1}$ for two’s-complement representations. Substituting Eq.~\eqref{eq:qj_binary} into Eqs.~\eqref{eq:Qk_cos}--\eqref{eq:Qk_sin} yields a decomposition of $e^{i\theta \hat Q_k^{(c)}}$ and $e^{i\theta \hat Q_k^{(s)}}$ into products of single-qubit $R_z$ rotations with angles
\begin{equation}
\phi^{(c)}_{j,r} = \theta\,\Delta_q\,w^{(c)}_{j,k}\,w_r^{\mathrm{bit}},
\qquad
\phi^{(s)}_{j,r} = \theta\,\Delta_q\,w^{(s)}_{j,k}\,w_r^{\mathrm{bit}},
\label{eq:angles}
\end{equation}
where $w^{(c)}_{j,k} = \tfrac{1}{\sqrt{N}}\cos(2\pi j k/N)$ and
$w^{(s)}_{j,k} = \tfrac{1}{\sqrt{N}}\sin(2\pi j k/N)$.
Because all $R_z$ rotations commute, the gates may be executed in parallel, resulting in circuits of constant depth. The corresponding circuits for the cosine and sine quadratures are shown in Fig.~\ref{fig:Qk_circuits}.

\subsection{Correlation-Function Estimation and Error Analysis}
\label{sec:Correlation-Function Estimation and Error Analysis}

The full correlator in Eq.~\eqref{eq:Qk_reconstruct} is obtained from two-time expectation values of the Hermitian quadratures $\hat Q_k^{(c)}$ and $\hat Q_k^{(s)}$. To access these quantities, we introduce four generating functions with independent source parameters coupled at times $t$ and $0$,
\begin{align}
F_c(\theta_1,\theta_2;t) &=
\big\langle e^{i\theta_1 \hat Q_k^{(c)}(t)} e^{i\theta_2 \hat Q_k^{(c)}(0)} \big\rangle, \\
F_s(\theta_1,\theta_2;t) &=
\big\langle e^{i\theta_1 \hat Q_k^{(s)}(t)} e^{i\theta_2 \hat Q_k^{(s)}(0)} \big\rangle, \\
F_{cs}(\theta_1,\theta_2;t) &=
\big\langle e^{i\theta_1 \hat Q_k^{(c)}(t)} e^{i\theta_2 \hat Q_k^{(s)}(0)} \big\rangle, \\
F_{sc}(\theta_1,\theta_2;t) &=
\big\langle e^{i\theta_1 \hat Q_k^{(s)}(t)} e^{i\theta_2 \hat Q_k^{(c)}(0)} \big\rangle .
\end{align}

\begin{figure}[t]
\centering
\footnotesize
\begin{quantikz}[row sep=0.40cm, column sep=0.28cm]
\lstick{\scriptsize Anc.} 
  & \ket{0} 
  & \gate{H} 
  & \ctrl{1} 
  & \qw  
  & \ctrl{1} 
  & \qw 
  & \gate{H} 
  & \meter{} \\
\lstick{\scriptsize Sys.} 
  & \ket{\psi_0} 
  & \qw
  & \gate{V_2}
  & \gate{U}
  & \gate{V_1}
  & \gate{U^\dagger}
  & \qw 
  & \qw
\end{quantikz}
\caption{Quantum workflow for evaluating the target quantity. The controlled operations are
$V_2 = e^{i\theta_2 \hat Q_k^{(c/s)}(0)}$ and
$V_1 = e^{i\theta_1 \hat Q_k^{(c/s)}(0)}$.
The time-shifted operator
$e^{i\theta_1 \hat Q_k^{(c/s)}(\tau)}$
is implemented in Heisenberg form as
$U^\dagger(\tau)\, e^{i\theta_1 \hat Q_k^{(c/s)}(0)}\, U(\tau)$.
After the final Hadamard gate, measurement of the ancilla in the $X$ basis yields
$\operatorname{Re}\langle V\rangle$.}
\label{fig:quantum_workflow}
\end{figure}
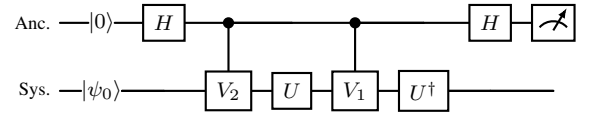

Each generating function may be estimated using a Hadamard-test-based circuit in which the exponentials of $\hat Q_k^{(c)}$ or $\hat Q_k^{(s)}$ are implemented at the appropriate time arguments. Time-shifted operators are realized in the Heisenberg picture via conjugation by the time-evolution operator. The full workflow is illustrated schematically in Fig.~\ref{fig:quantum_workflow}.
The desired correlators are extracted from these generating functions by evaluating mixed derivatives at the origin. For example,
\begin{equation}
-\partial_{\theta_1}\partial_{\theta_2}
F_c(\theta_1,\theta_2;t)
\big|_{\theta_1=\theta_2=0}
=
\langle \hat Q_k^{(c)}(t)\hat Q_k^{(c)}(0) \rangle .
\end{equation}
Analogous expressions obtained from $F_s$, $F_{cs}$, and $F_{sc}$ provide the remaining terms required for Eq.~\eqref{eq:Qk_reconstruct}.

In practice, the mixed derivatives are approximated by finite-difference estimators. To eliminate contributions from single-source moments, we employ the rectangle-rule estimator
\begin{equation}
\widehat C_{\mathrm{rect}}(t;h)
=
-\frac{
F(h,h;t) - F(h,0;t) - F(0,h;t) + F(0,0;t)
}{h^2},
\end{equation}
which isolates the mixed derivative exactly at leading order. A Taylor expansion shows that the associated systematic bias scales linearly with the step size,
\begin{equation}
\mathrm{bias}_{\mathrm{rect}}(t;h) = \mathcal{O}(h\,Q_{\max}^3),
\end{equation}
where $Q_{\max}$ denotes the maximum encoded displacement. Higher-order finite-difference schemes or Richardson extrapolation may be used to suppress the leading bias at the cost of additional circuit evaluations. Detailed derivations and error bounds are provided in Appendix~\ref{app:fd-error}.

In addition to systematic bias, statistical uncertainty arising from finite sampling is amplified by the finite-difference prefactor and scales as
\begin{equation}
\epsilon_{\mathrm{stat}} \sim \frac{1}{h^2\sqrt{N_{\mathrm{shots}}}}.
\end{equation}
The finite-difference step size $h$ must therefore be chosen to balance systematic and statistical errors.

\section{Resource Estimation}\label{sec:Resource_Estimation} 
In this section, we estimate the quantum resources required to implement the simulation protocol developed in Sections~\ref{sec:Mathematical_Formulation}--\ref{sec:Fourier_Mode_Displacement_Operator}. 
All estimates are based on the first-quantized Hamiltonian formulation, the symmetric Trotterization strategy described in Section~\ref{Unitary_for_hamiltonian_evolution}, and the explicit circuit constructions for the kinetic and potential evolution operators.
Throughout this section, the focus is on resource accounting at the level of executable workflows, including unitary evolution, measurement overhead, and correlator reconstruction, rather than on isolated circuit primitives.
Our emphasis is on asymptotic scaling and order-of-magnitude estimates relevant to fault-tolerant quantum computing, rather than near-term feasibility.

\subsection{Qubit Requirements and Trotter Step Count}

Each lattice displacement $q_j$ is represented on a uniform $b$-qubit register, corresponding to a discrete grid
$q_j \in [-q_{\max}, q_{\max}]$ with spacing $\Delta q = 2q_{\max}/2^{b}$. The system register therefore requires
\begin{equation}
Q_{\mathrm{sys}} = N b
\end{equation}
logical qubits.

Implementation of the potential operator requires temporary ancilla registers to store the relative displacements
\begin{equation}
\Delta_j = q_{j+1} - q_j ,
\end{equation}
which are computed via reversible subtraction and uncomputed after the controlled phase rotations are applied. To enable parallel execution of non-overlapping interactions, the lattice is partitioned into even and odd bond layers, requiring at most
\begin{equation}
Q_{\mathrm{anc}} = \frac{N}{2}\, b
\end{equation}
ancilla qubits. An additional $\mathcal{O}(1)$ ancilla qubit is required for the Hadamard-test-based correlator estimation. The total logical-qubit requirement is therefore
\begin{equation}
Q_{\mathrm{tot}} = \frac{3}{2} N b + \mathcal{O}(1).
\end{equation}

Time evolution is implemented using the symmetric second-order Trotter--Suzuki decomposition described in Section~\ref{Unitary_for_hamiltonian_evolution}. The local error per Trotter step scales as
\begin{equation}
\epsilon_{\mathrm{step}}
=
\mathcal{O}\!\left(
\Delta t^{3}
\big(
[H_{\mathrm{kin}},[H_{\mathrm{kin}},H_{\mathrm{pot}}]]
+
[H_{\mathrm{pot}},[H_{\mathrm{pot}},H_{\mathrm{kin}}]]
\big)
\right),
\end{equation}
so that after $n=t/\Delta t$ steps the accumulated error scales as $\mathcal{O}(t\,\Delta t^{2})$, up to operator-norm prefactors that increase with the anharmonicity parameter $\beta$. Fixing a target simulation accuracy $\varepsilon$ therefore implies
\begin{equation}
n = \mathcal{O}\!\left(\frac{t^{3/2}}{\sqrt{\varepsilon}}\right).
\end{equation}

\subsection{Gate and Depth Complexity of the Execution Workflow}

Each kinetic half-step is implemented by applying a quantum Fourier transform (QFT) to each site register, followed by diagonal phase rotations and an inverse QFT. 
For a $b$-qubit register, an exact QFT requires $\mathcal{O}(b^{2})$ two-qubit gates and has depth $\mathcal{O}(b)$ \cite{nielsen2010quantum}. 
Executing this procedure on all $N$ sites in parallel yields a kinetic-layer gate count
\begin{equation}
G_{\mathrm{kin}} = \mathcal{O}(N b^{2}),
\end{equation}
with depth $\mathcal{O}(b)$.

The potential layer consists of reversible subtraction, controlled quadratic and quartic phase rotations, and uncomputation. 
Reversible subtraction on $b$-bit registers requires $\mathcal{O}(b)$ Toffoli-class gates, while synthesis of the quartic phase $\exp[-i(\beta/4)\Delta_j^{4}\Delta t/\hbar]$ involves reversible multiplication with cost $\mathcal{O}(b^{2})$ \cite{PhysRevA.54.147}.
Summed over all bonds, the potential-layer gate count also scales as
\begin{equation}
G_{\mathrm{pot}} = \mathcal{O}(N b^{2}),
\end{equation}
with depth $\mathcal{O}(b)$ under even--odd layer parallelization.

\begin{figure}
\centering
\includegraphics[scale=0.5]{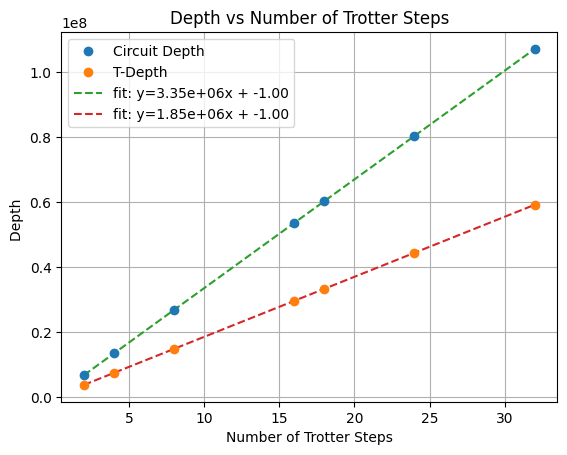}
\caption{Circuit depth versus number of Trotter steps for the simulation of the quantum $\beta$--FPUT model, obtained using Qiskit with circuit optimization level 3. The depth increases linearly with the number of time steps, but with a large constant prefactor.}
\label{fig:Circuit_Depth_Scalling_from_Simulation}
\end{figure}

Combining both contributions, the total gate count per Trotter step is
\begin{equation}
G_{\mathrm{step}} = \mathcal{O}(N b^{2}),
\end{equation}
and the circuit depth per step is $\mathcal{O}(b)$. 
Figure~\ref{fig:Circuit_Depth_Scalling_from_Simulation} illustrates the depth scaling obtained from an explicit circuit construction implemented in Qiskit, showing linear growth with the number of Trotter steps and a large constant prefactor.

Extraction of the displacement correlator $\langle \hat Q_k(t)\hat Q_k(0)\rangle$ follows the Hadamard-test workflow described in Section~\ref{sec:Fourier_Mode_Displacement_Operator}. Four independent generating functions are evaluated, and mixed derivatives are approximated using finite differences. Sampling $N_\tau$ time lags with $M$ circuit repetitions per expectation value requires
\begin{equation}
N_{\mathrm{exec}} = 4 N_\tau M
\end{equation}
circuit executions. 
Each execution involves forward and backward real-time evolution for $n(\tau)$ Trotter steps, so the overall computational cost scales linearly with the simulated time and with the number of sampled time points.

\section{Conclusion and Outlook}\label{sec:Conclusion_Outlook}

We have presented a first-quantized digital quantum-simulation framework for the quantum $\beta$--FPUT lattice. Starting from the real-space Hamiltonian, we derived its quantum formulation, constructed Trotterized circuit blocks for kinetic and anharmonic potential evolution, and specified a measurement workflow for mode-resolved displacement correlators relevant to phonon relaxation and anomalous transport.

A central algorithmic point is how Fourier-mode displacement observables are handled in a digital setting. The operator $\hat Q_k$ is non-Hermitian, so it cannot be exponentiated directly as a unitary. By decomposing it into commuting Hermitian cosine and sine quadratures, we obtain implementable unitaries whose circuits reduce to parallel single-qubit $R_z$ rotations under the displacement-grid encoding. This makes the correlator $\langle \hat Q_k(t)\hat Q_k(0)\rangle$ accessible through Hadamard-test evaluations of generating functions combined with finite-difference estimators, while retaining a clean separation between unitary time evolution and observable extraction.

We also analyzed the resource requirements of the full workflow in a fault-tolerant setting. For an $N$-site chain encoded with $b$ qubits per site, the logical-qubit requirement scales as $Q_{\mathrm{tot}}=\tfrac{3}{2}Nb+\mathcal{O}(1)$ under parallel-bond execution, and the gate complexity per Trotter step scales as $\mathcal{O}(Nb^{2})$. Together with the depth prefactors implied by reversible arithmetic, basis changes, and correlator estimation overhead, these results frame the method as a fault-tolerant roadmap rather than a near-term approach.

Several next steps are both technically concrete and algorithmically meaningful. First, benchmarking small system sizes against exact diagonalization and controlled semiclassical limits would validate the combined effects of discretization, Trotterization, and correlator reconstruction, and help calibrate the constant-factor overheads that dominate practical depth. Second, an end-to-end estimator study can be used to choose finite-difference step sizes and sampling budgets that balance systematic bias against shot-noise amplification. Third, extending the observable set from displacement correlators to momentum and energy-current correlators would tighten the connection to transport quantities and Green--Kubo-type formulations. Finally, developing state-preparation strategies for low-temperature and thermally distributed initial states would move the present framework toward complete workflows for quantum transport in nonlinear lattice systems, including regimes where symmetry assumptions do not suppress mixed quadrature contributions.

\section*{Acknowledgements}

The author acknowledges the use of \textit{ChatGPT} (OpenAI, GPT-5 model) as a writing assistant in the preparation of this article. 
The tool was employed to improve the clarity, structure, and flow of the text. 
All scientific content, mathematical derivations, and interpretations were independently verified and finalized by the authors.
The authors are also grateful for the insightful discussion and comments from Rafael Augusto Couceiro Correa, Marek Kowalik and Walden Killick.
This work was supported by the MSCA Cofund QuanG programme under Grant No. 101081458, funded by the European Union. K.N.M.S. acknowledges support from this programme. Views and opinions expressed are, however, those of the author(s) only and do not necessarily reflect those of the European Union or the granting authority.
Neither the European Union nor the granting authority can be held responsible for them.
\begin{center}
    \includegraphics[width=0.25\linewidth]{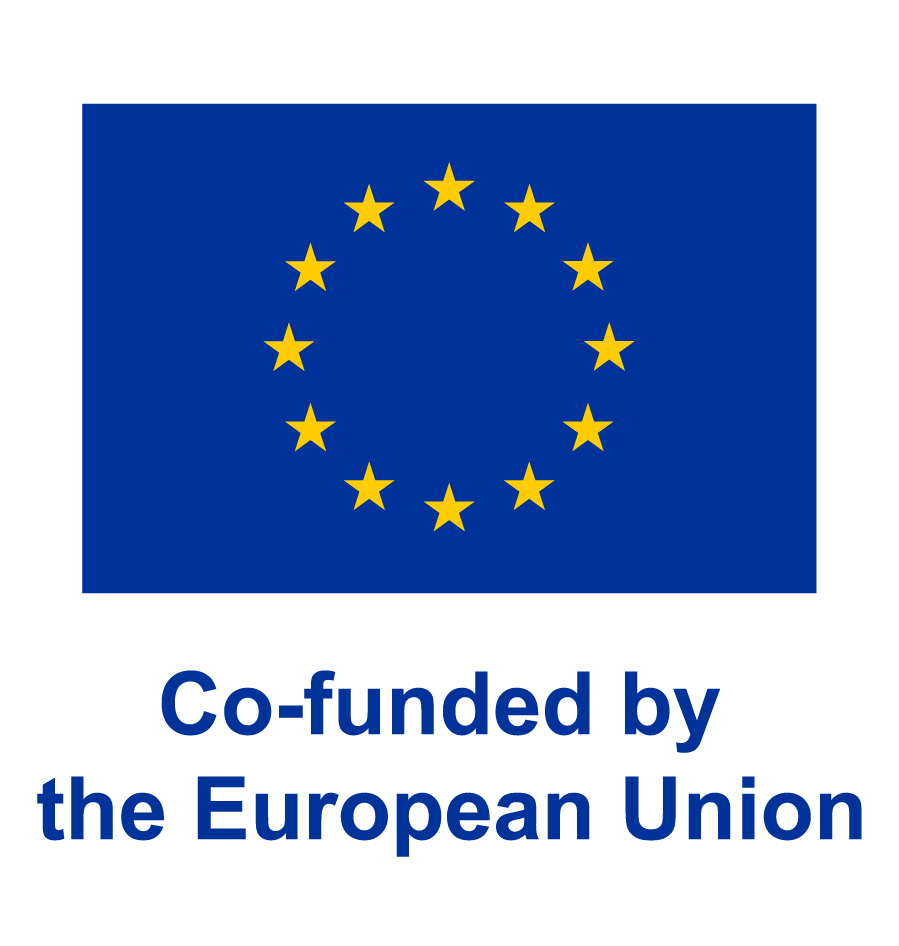}
\end{center}
\section*{Author Contribution}
K.N.M.S. conceived the project, carried out the majority of the technical work, and wrote the initial draft of the manuscript, which was revised under the supervision of P.L. P.L. provided overall scientific direction and contributed to writing and shaping the manuscript. J.M.A.H. and J.V.V. provided feedback during manuscript development, with J.V.V. additionally advising on the positioning of the work in the context of broader industrial applications. The manuscript was proof read by all other authors.

\bibliographystyle{unsrt}   % or another style: plain, IEEEtran, etc.
\bibliography{references}   % this matches your .bib filename, without the .bib extension

\appendix
\section{Proof that $\big[\hat Q_k^{(c)},\,\hat Q_k^{(s)}\big]=0$}

\paragraph{Assumptions and definitions.}
Consider a one-dimensional chain of $N$ lattice sites with canonical position
operators $\{\hat q_j\}_{j=0}^{N-1}$, each Hermitian with units of length and
satisfying
\begin{equation}
[\hat q_j,\hat q_\ell]=0\qquad\forall\,j,\ell\in\{0,\dots,N-1\}.
\end{equation}
For a given wave number $k\in\{0,1,\dots,N-1\}$ define the (dimensionless) phases
$\varphi_{jk}=2\pi jk/N$ and the Hermitian Fourier quadratures
\begin{equation}
\hat Q_k^{(c)}=\frac{1}{\sqrt{N}}\sum_{j=0}^{N-1}\hat q_j\cos\varphi_{jk},
\qquad
\hat Q_k^{(s)}=\frac{1}{\sqrt{N}}\sum_{j=0}^{N-1}\hat q_j\sin\varphi_{jk}.
\end{equation}

\paragraph{Claim.} The cosine and sine quadratures commute:
\begin{equation}
\big[\hat Q_k^{(c)},\,\hat Q_k^{(s)}\big]=0\qquad\text{for all }k.
\end{equation}

\paragraph{Proof}
Both $\hat Q_k^{(c)}$ and $\hat Q_k^{(s)}$ are \emph{real} linear combinations of the mutually commuting set
$\{\hat q_j\}$. Any two real linear combinations of a commuting family commute. Hence
$[\hat Q_k^{(c)},\hat Q_k^{(s)}]=0$.

\paragraph{Corollary (exact factorization of exponentials).}
For any real angles $\alpha,\beta$,
\begin{equation}
e^{i\alpha \hat Q_k^{(c)}+i\beta \hat Q_k^{(s)}}
= e^{i\alpha \hat Q_k^{(c)}}\,e^{i\beta \hat Q_k^{(s)}}
= e^{i\beta \hat Q_k^{(s)}}\,e^{i\alpha \hat Q_k^{(c)}},
\end{equation}
i.e., any Trotter splitting between the cosine and sine generators is \emph{exact}.

\section{Error Analysis of the Finite-Difference Estimator}
\label{app:fd-error}

To quantify the systematic error of the rectangle-rule estimator, we expand the
generating function
\begin{equation}
F(\theta_1,\theta_2;t)
= \big\langle e^{i\theta_1 \hat Q_k(t)} e^{i\theta_2 \hat Q_k(0)} \big\rangle
\end{equation}
in a multivariate Taylor series about $(\theta_1,\theta_2)=(0,0)$.
Writing $F_{ab}=\partial_{\theta_1}^a \partial_{\theta_2}^b F(0,0;t)$,
the series reads
\begin{align}
F(h,0) &= F_{00} + h F_{10} + \tfrac{h^2}{2}F_{20} + \tfrac{h^3}{6}F_{30} + O(h^4), \\
F(0,h) &= F_{00} + h F_{01} + \tfrac{h^2}{2}F_{02} + \tfrac{h^3}{6}F_{03} + O(h^4), \\
F(h,h) &= F_{00} + h(F_{10}+F_{01})
+ \tfrac{h^2}{2}(F_{20}+2F_{11}+F_{02}) \nonumber \\
&\qquad + \tfrac{h^3}{6}(F_{30}+3F_{21}+3F_{12}+F_{03}) + O(h^4),
\end{align}
where $F_{00}=F(0,0;t)=1$.

The rectangle combination appearing in the estimator is
\begin{equation}
S(h) := F(h,h)-F(h,0)-F(0,h)+F(0,0).
\end{equation}
Substituting the expansions gives
\begin{equation}
S(h) = h^2 F_{11} + \tfrac{h^3}{2}(F_{21}+F_{12}) + O(h^4).
\end{equation}
Thus the rectangle estimator
\begin{equation}
\widehat C_{\mathrm{rect}}(t;h) = -\frac{S(h)}{h^2}
= -F_{11} - \tfrac{h}{2}(F_{21}+F_{12}) + O(h^2)
\end{equation}
approximates the desired correlator $C(t) = \langle \hat Q_k(t)\hat Q_k(0)\rangle = -F_{11}$ up to a bias term.

The bias is therefore
\begin{equation}
\mathrm{bias}_{\mathrm{rect}}(t;h)
= \widehat C_{\mathrm{rect}}(t;h)-C(t)
= -\tfrac{h}{2}\,(F_{21}+F_{12}) + O(h^2).
\end{equation}
Using the definitions of the derivatives,
\begin{equation}
F_{21} = -\,i\langle \hat Q_k(t)^2 \hat Q_k(0)\rangle,
\qquad
F_{12} = -\,i\langle \hat Q_k(t)\hat Q_k(0)^2\rangle ,
\end{equation}
we can write the leading bias explicitly as
\begin{equation}
\mathrm{bias}_{\mathrm{rect}}(t;h)
= \tfrac{i h}{2}\,\langle \hat Q_k(t)^2 \hat Q_k(0) + \hat Q_k(t)\hat Q_k(0)^2 \rangle + O(h^2).
\end{equation}

If the operator norm of $\hat Q_k$ is bounded by $Q_{\max}$ (as is the case under $b$-bit encoding with dynamic range),
then $\| \hat Q_k(t)^2 \hat Q_k(0)\|,\,\|\hat Q_k(t)\hat Q_k(0)^2\|\le Q_{\max}^3$, yielding the bound
\begin{equation}
\big|\mathrm{bias}_{\mathrm{rect}}(t;h)\big|\;\le\; h\,Q_{\max}^3 + O(h^2).
\end{equation}
This proves that the rectangle estimator is first-order accurate in $h$.

Improved accuracy is possible. A central mixed-difference scheme using points $(\pm h,\pm h)$ cancels
the cubic terms and reduces the bias to $O(h^2 Q_{\max}^4)$. Alternatively, Richardson extrapolation
between step sizes $h$ and $h/2$ achieves the same $O(h^2)$ scaling without requiring negative displacements.

\section{Illustrative Quantum Hardware Roadmaps}
\label{app:hardware_roadmap}

To provide broader context for the fault-tolerant resource estimates discussed in Section~\ref{sec:Resource_Estimation}, Table~\ref{tab:roadmap} summarizes publicly stated quantum hardware roadmap targets from selected industrial and academic efforts. The entries are intended solely as illustrative reference points. No claim is made regarding the suitability of any specific platform or timeline for implementing the algorithmic framework developed in this work.

% \begin{table}[h]
% \centering
% \footnotesize
% \renewcommand{\arraystretch}{1.15}
% \setlength{\tabcolsep}{4pt}
% \begin{tabular}{p{2.2cm} p{3.0cm} p{4.2cm} p{2.6cm}}
% \hline
% \textbf{Organization} & \textbf{Platform} & \textbf{Publicly stated roadmap target(s)} & \textbf{Indicative timeframe} \\
% \hline
% IBM &
% Superconducting qubits &
% $\sim$2000 physical qubits and $\sim10^9$ gate operations &
% Early 2030s \\
% \hline
% IonQ &
% Trapped ions &
% $\sim$2 million physical qubits ($\sim$80{,}000 logical qubits) &
% By late 2020s \\
% \hline
% QuEra &
% Neutral atoms &
% $\sim$10{,}000 physical qubits ($\sim$100 logical qubits) &
% Mid 2020s \\
% \hline
% Xanadu &
% Photonic qubits &
% $\sim$100{,}000 physical qubits ($\sim$500--1000 logical qubits) &
% Late 2020s \\
% \hline
% \end{tabular}
% \caption{Illustrative quantum hardware roadmap targets as stated in public communications. Values are approximate and subject to change. The table is included for context only and does not imply endorsement, feasibility, or suitability for the simulation protocol introduced in this work.}
% \label{tab:roadmap}
% \end{table}

\begin{table}[h]
\centering
\footnotesize
\renewcommand{\arraystretch}{1.15}
\resizebox{\linewidth}{!}{
\begin{tabular}{l l p{5.5cm} l}
\hline
\textbf{Organization} &
\textbf{Platform} &
\textbf{Publicly stated roadmap target(s)} &
\textbf{Timeframe} \\
\hline
IBM & Superconducting qubits & $\sim$2000 physical qubits and $\sim10^9$ gate operations & Early 2030s \\
IonQ & Trapped ions & $\sim$2M physical qubits ($\sim$80{,}000 logical qubits) & Late 2020s \\
QuEra & Neutral atoms & $\sim$10{,}000 physical qubits ($\sim$100 logical qubits) & Mid 2020s \\
Xanadu & Photonic qubits & $\sim$100{,}000 physical qubits ($\sim$500--1000 logical qubits) & Late 2020s \\
\hline
\end{tabular}
}
\caption{Illustrative quantum hardware roadmap targets based on publicly stated information.}
\label{tab:roadmap}
\end{table}

\end{document}